\documentclass[aip, apl, superscriptaddress, reprint]{revtex4-1}

\usepackage{siunitx}
\usepackage{chemformula}
\usepackage{graphicx}
\usepackage{hyperref}
\usepackage{verbatim}

\newcommand{\mr}[1]{\mathrm{#1}}

\newcommand{\cf}[0]{cf.\ }
\newcommand{\ie}[0]{i.e.\ }
\newcommand{\egg}[0]{e.g.\ }
\newcommand{\Fref}[1]{Figure~\ref{fig:#1}}
\newcommand{\fref}[1]{Fig.~\ref{fig:#1}}

\newcommand{\eref}[1]{Eq.~(\ref{eq:#1})}
\newcommand{\Cref}[1]{Chapter~\ref{chap:#1}}
\newcommand{\cref}[1]{Ch.~\ref{chap:#1}}

\renewcommand{\vec}[1]{{\mathrm{\mathbf{#1}}}}

\begin{document}

\title{Non-local magnon-based transport in yttrium iron garnet/platinum heterostructures at high temperatures}

\author{Richard Schlitz}
\email{richard.schlitz@tu-dresden.de}
\affiliation{Institut f{\"u}r Festk{\"o}rper- und Materialphysik and Würzburg-Dresden Cluster of Excellence ct.qmat, Technische Universit{\"a}t Dresden, 01062 Dresden, Germany}
\author{Sergey Granovsky}
\affiliation{Institut f{\"u}r Festk{\"o}rper- und Materialphysik and Würzburg-Dresden Cluster of Excellence ct.qmat, Technische Universit{\"a}t Dresden, 01062 Dresden, Germany}
\affiliation{Faculty of Physics, M.\ V.\ Lomonossow Moscow State University, Moscow 119991, Russia}
\author{Sebastian T. B. Goennenwein}
\affiliation{Institut f{\"u}r Festk{\"o}rper- und Materialphysik and Würzburg-Dresden Cluster of Excellence ct.qmat, Technische Universit{\"a}t Dresden, 01062 Dresden, Germany}
\affiliation{Department of Physics, University of Konstanz, 78457 Konstanz, Germany}

\date{\today}

\begin{abstract}
    The spin Hall effect in a heavy metal thin film allows to probe the magnetic properties of an adjacent magnetic insulator via magnetotransport measurements.
    Here, we investigate the magnetoresistive response of yttrium iron garnet/platinum heterostructures from room temperature to beyond the Curie temperature $T_\mr{C, YIG} \approx \SI{560}{\kelvin}$ of the ferrimagnetic insulator.
    %In contrast to previous studies where a high current is used to locally heat the device, we infer the high temperature response by steady state heating and using only a small excitation current, reducing possible artifacts. 
	We find that the amplitude of the (local) spin Hall magnetoresistance decreases monotonically from \SI{300}{\kelvin} towards $T_\mr{C}$, mimicking the evolution of the saturation magnetization of yttrium iron garnet. 
    Interestingly, the spin Hall magnetoresistance vanishes around \SI{500}{\kelvin}, well below $T_\mr{C}$, which we attribute to the formation of a parasitic interface layer by interdiffusion.
Around room temperature the non-local magnon-mediated magnetoresistance exhibits a power law scaling $T^{\alpha}$ with $\alpha = 3/2$, as already reported. 
The exponent decreases gradually to $\alpha \sim 1/2$ at around \SI{420}{\kelvin}, before the non-local magnetoresistance vanishes rapidly at a similar temperature as the spin Hall magnetoresistance.
   We attribute the reduced $\alpha$ at high temperatures to the increasing thermal magnon population which leads to enhanced scattering of the non-equilibrium magnon population and a reduced magnon diffusion length.
   Finally, we find a magnetic field independent offset voltage in the non-local signal for $T > \SI{470}{\kelvin}$ which we associate with electronic leakage currents through the normally insulating yttrium iron garnet film.
    Indeed, this non-local offset voltage is thermally activated with an energy close to the band gap.
\end{abstract}

\maketitle

\section{Introduction}

Magnon spintronics is emerging as a new subfield of spintronics, in which wave computing schemes based on magnons are one of the major goals.\cite{Chumak2015}
To that end, the transport properties of angular momentum via magnons needs to be well characterized in different materials, and schemes for the electric generation and detection of magnon currents must be developed.
One way to access the magnetic properties and the transport of magnons in magnetic insulators is provided by spin current based experiments in heterostructures of a magnetic insulator (MI) and a heavy metal (HM) using electrical detection.\cite{Aqeel2015, Chen2013, Althammer2013, Nakayama2013, Cornelissen2015, Goennenwein2015, Zhang2012, Zhang2012a, Ganzhorn2016, Hoogeboom2017, Lebrun2018}
In such experiments, the spin Hall effect in the HM layer is used to generate a spin accumulation at the MI/HM interface by sourcing an electrical current.\cite{Hirsch1999, Chen2013, Zhang2012}
This spin accumulation can then interact locally with the magnetic moments of the MI at the interface, leading to a dissipation of angular momentum, \egg via the spin transfer torque.\cite{Slonczewski1996, Chen2013}
In combination with the inverse spin Hall effect, this dissipation leads to the so-called spin Hall magnetoresistance (SMR).\cite{Nakayama2013, Althammer2013,Chen2013}
In a complementary process, the interfacial spin accumulation excites magnons in the MI, which then diffuse through the MI and can be detected by a second, electrically insulated HM wire in close vicinity.
This effect is usually referred to as the non-local magnon-mediated magnetoresistance (MMR).\cite{Cornelissen2015, Goennenwein2015, Zhang2012, Zhang2012a}

Consequently, investigations of the SMR and the MMR as a function of temperature allow to infer the temperature dependence of the interfacial spin transparency, the spin diffusion length, the spin Hall angle, the magnon generation efficiency and the magnon diffusion length.\cite{Goennenwein2015, Marmion2014, Cornelissen2015, Meyer2015, Isasa2015, Wang2015, Uchida2015, Cornelissen2016, Thiery2018}
To date, the exact behavior of the SMR and MMR around the Curie temperature of the MI is not well understood.
There have been several reports of a large SMR or MMR in heterostructures of a HM and a disordered or paramagnetic magnetic insulator.\cite{Oyanagi2019, Schlitz2018a, Oyanagi2020, Miura2020, Wesenberg2017}
However, other experiments showed that in the absence magnetic ordering no MMR can be observed.\cite{Schlitz2019a, Gomez-Perez2020}
From the theory side, microscopic models to describe the temperature dependence of the SMR have been put forward and good agreement with experiments was found for different material systems.\cite{Zhang2019, Gomez-Perez2020a, Velez2019}
In particular, the SMR is expected to scale (in leading order) with the square of the saturation magnetization as found experimentally.\cite{Zhang2019, Wang2015, Uchida2015, Ohnuma2014}
Theoretical studies of the MMR showed that it should depend on temperature with a power law $T^\alpha$, where the magnon generation gives rise to a leading expontent of $\alpha^{em} = 3/2$ and the magnon detection to an additional exponent of $\alpha^{me} = 1$.\cite{Zhang2012a}
This was verified experimentally, where a power of $\alpha = 5/2$ was found for low temperatures $T \lesssim \SI{100}{\kelvin}$, while for higher temperatures up to room temperature $\alpha = 3/2$ was reported.\cite{Goennenwein2015}
However, at high temperatures the thermal magnon population and corresponding magnon scattering processes increase significantly, so that one might expect that the magnon diffusion incurs an additional temperature dependence.
So far, such non-local magnetotransport experiments have only been carried out at or below room temperature, where a weak dependence of the spin diffusion length on temperature was found. \cite{Cornelissen2016}
Thus, this expectation was never experimentally tested for higher temperatures.

%\rs{missing: olivier results towards higher temperatures using pulsed measurements?}

In this article, we report measurements of the SMR and the MMR in heterostructures of \ch{Y3Fe5O12} (YIG) and Pt up to above the Curie temperature of YIG.
We use uniform heating of the entire sample in combination with rotations of the magnetic field in the sample plane to extract the amplitudes of the SMR and MMR.
This approach minimizes detrimental heat gradients directly below the investigated metallic structures, which often are introduced by high currents used in Joule heating schemes.
We critically compare the observed temeprature scaling of the SMR and MMR amplitudes to theoretical expectations, and in particular address the influence of increased magnon scattering and of annealing (intermixing) arising for high temepratures.
%Using this appraoch, we find an unusual scaling in the SMR, where a direct proportionality to the saturation magnetization, contrary to previous reports.

\section{Methods}

We used a commercially available, \SI{150}{\nano\meter} thick YIG film deposited on a (111)-oriented \ch{Gd3Ga5O12} (GGG) substrate via liquid phase epitaxy to prepare our YIG/Pt heterostructures.
To clean the YIG film and remove potential organic surface contaminations, the sample was first etched in a Piranha solution (\ch{H2SO4:H2O2} in a 1:1 volumetric ratio) and subsequently annealed in the ultra high vacuum of the deposition chamber at \SI{200}{\celsius} for \SI{1}{\hour}.
A platinum film with a thickness of $t_\mr{Pt} = \SI{3}{\nano\meter}$ was then sputter deposited at room temperature.
Using optical lithography and \ch{Ar} ion milling the Pt film was patterned into Hall bars and parallel (electrically insulated) wires for the non-local transport experiments.
The Hall bars have a width of $w = \SI{50}{\micro\meter}$ and a length of $l = \SI{400}{\micro\meter}$ between the contacts used for measuring the longitudinal voltage (\cf \fref{f1}a).
The non-local devices have an edge to edge separation of $d_\mr{NL} \sim \SI{1}{\micro\meter}$ between the two Pt wires.
For the measurements discussed below, Pt wires with a length and width of $l_\mr{NL} \sim \SI{100}{\micro\meter}$ and $w_\mr{NL} \sim \SI{2}{\micro\meter}$ were used (\cf \fref{f1}b).

To record the local magnetoresistive response (\ie the SMR) on the Hall bars, we drive a DC current of $I = \SI{100}{\uA}$ along the Hall bar using a Keithley 2450 sourcemeter, and detect the longitudinal $V_{\ell,\mathrm{raw}}$ and transverse $V_\mr{t,raw}$ voltage drop using two Keithley 2182 nanovoltmeters (\cf \fref{f1}a). 
For the non-local response, again a current $I = \SI{100}{\micro\ampere}$ is sent through one of the two Pt wires using a sourcemeter and the non-local voltage $V_\mr{nl,raw}$ is recorded by an independent Keithley 2182 nanovoltmeters on the second Pt wire.
In order to remove spurious thermoelectric effects, to increase the sensi\-tivity and to single out the resistive (ohmic) response, we employ a current reversal technique and obtain\cite{Goennenwein2015, Avci2015, Schreier2013} 
\begin{equation}
	V_i = \frac{V_{i,\mr{raw}}(+I) - V_{i,\mr{raw}}(-I)}{2}.
	\label{eq:asym}
\end{equation}
The longitudinal and transverse resitivities can then be determined from the measured voltages as $\rho_\ell = V_\ell w t_\mr{Pt} / l$ and $\rho_\mr{t} = V_\mr{t} t_\mr{Pt}$, respectively, using the dimensions of the Pt structures quoted above.
A magnetic field $\mu_0 H \approx \SI{70}{\milli\tesla}$ is applied using a diametrically magnetized cylindrical Halbach array.\cite{Halbach1980}
By rotating this Halbach array (the angle of rotation is denoted as $\alpha$, see \fref{f1}a and b), we measure the magnetotransport response as a function of the magnetic field orientation in the YIG/Pt interface plane.

The resistivity at temperatures $T>\SI{300}{\kelvin}$ is measured using a custom-built vacuum oven.
In this oven, the sample (mounted on a ceramic chip carrier) is attached to a copper heat sink. The latter can be heated via a cartridge heater inserted into the heat sink.
The sample is glued to the chip carrier with silver paint and contacted via aluminum wedge bonding.
Electrical contact to the chip carrier is established with high temperature compatible spring loaded contact pins.
The temperature is measured using a Pt100 resistance thermometer directly attached to the copper heatsink.

Finally, to mitigate the impact of fluctuations and drifts of the oven temperature on the magnetotransport measurement, we here focus on the transverse resistivity $\rho_\mr{t}$ to determine the magnetoresistance ratio $\Delta \rho/\rho_0$ as introduced in Ref.~\cite{Aqeel2015} by Aqeel et al..
This is motivated by the fact that the expected SMR signature for rotations of the magnetic field in the sample plane (assuming a saturated magnetization) is given by\cite{Chen2013, Nakayama2013, Althammer2013}
\begin{align}
    \rho_\ell &= \rho_0 - \Delta \rho \sin^2(\alpha)\\
    \rho_\mr{t} &= \Delta \rho \sin(\alpha)\cos(\alpha).\label{eq:smrt}
\end{align}
$\Delta \rho$ is the amplitude of the resistivity modulation and $\rho_0$ is the magnetic field orientation independent resistivity of the Pt film.
Consequently, $\Delta \rho$ can be very sensitively extracted from the transverse resistivity, without a detrimental impact of $\rho_0(T)$ changes, while $\rho_0$ is obtained from the (average) longitudinal resistivity. 
For the non-local voltage arising from the MMR one would expect 
\begin{equation}
    V_\mr{nl} =  \Delta V_\mr{nl} \sin^2(\alpha),\label{eq:nl}
\end{equation}
where the amplitude $\Delta V_\mr{nl} \le 0$ for the contact polarity used here.\cite{Goennenwein2015, Cornelissen2015, Zhang2012, Zhang2012a}

In addition to magnetotransport, we also performed high temperature magnetization measurements in an Oxford Instruments MagLab vibrating sample magnetometer using the high-temperature insert provided by the company. 
The sample hereby was cemented in the ceramic sample stick, heated to 630K and subsequently held at this temperature for 2 hours to minimize thermal gradients in the insert. 
Then, the temperature was swept down with \SI{1}{\kelvin\per\minute} and the magnetic moment vs.\ temperature dependence was recorded while a constant magnetic field of $\mu_0 H = \SI{7}{\milli\tesla}$ was applied in the thin film plane, fully saturating the YIG magnetization. 
To ensure the correct temperature reading, the same sequence was performed also with a reference thermocouple cemented in the insert at the sample position.

\section{Results and Discussion}

The normalized transverse resistivty -- which directly yields the SMR ampltiude $\Delta \rho/\rho_0$ as discussed above -- is shown in \fref{f1}(c) and (e) for selected temperatures below and above \SI{450}{\kelvin}, respectively. 
Please note, that drifts in $\rho_\ell$ (used to determine $\rho_0$ by averaging over all angles) introduce slight errors in the SMR amplitude.
However, since these drifts are well below one percent of $\rho_0$, the error introduced to the magnetoresistance ratio is negligible.
As expected for the SMR, a $\sin(\alpha)\cos(\alpha)$ (\cf \eref{smrt}) is visible in the $\rho_t$ data.
While below \SI{450}{\kelvin}, only a small decrease of the modulation amplitude $\Delta \rho$ (\ie the SMR) is evident, a sharp drop above \SI{480}{\kelvin} is evident from \fref{f1} and no clear magnetoresistance signature can be picked up beyond noise and drifts above \SI{540}{\kelvin}.

The $V_\mr{nl}$ data, shown in \fref{f1}(d) and (f) for the same temperatures as the transverse resistivity, reproduce this signature.
However, for the higher temperatures (panel (f)), the modulation is only visible up to \SI{510}{\kelvin} and an additional (field independent) offset signal is superimposed onto the modulation.
This offset increases significantly with increasing temperature and is the fingerprint of a finite electrical conductivity of the YIG film, giving rise to finite electric leakage currents, which will be addressed in more detail below (\cf \fref{f4}).\cite{Gomez-Perez2020, Thiery2018, Thiery2018a}

\onecolumngrid
%\rule{500pt}{1pt}

\begin{figure}[h!]
    \includegraphics{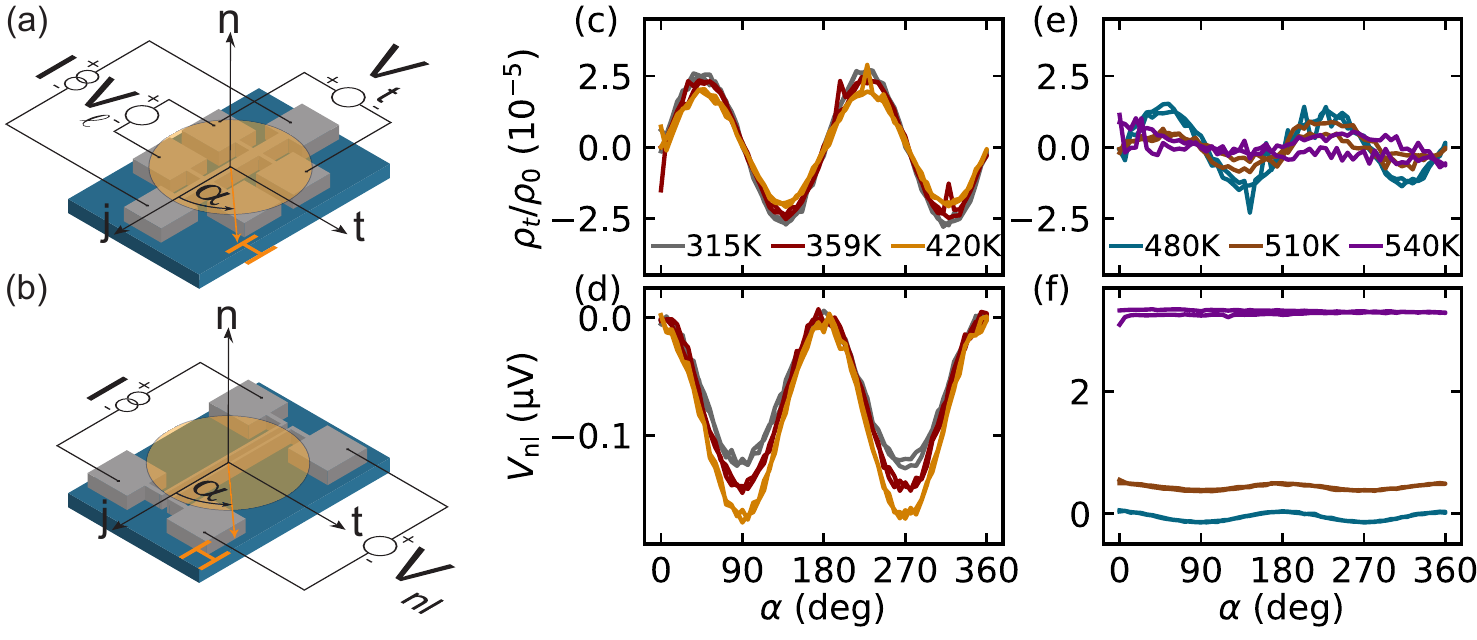}
    \caption{\label{fig:f1} 
	(a) The Hall bar used for measuring the local magnetotransport response (SMR).
        A current is applied along the $\vec{j}$ direction and the  longitudinal $V_\mr{\ell}$ and the transverse $V_\mr{t}$ voltage drop are measured along the current and along the transverse $\vec{t}$ direction, respectively.
	(b) In the non-local device geometry, the current is applied in one Pt wire, while the non-local voltage $V_\mr{nl}$ is measured on a second, electrically insulated Pt wire.
        The magnetic field is rotated in the sample plane for both measurement types.        
	Local (panels (c) and (e)) and non-local (panels (d) and (f)) magnetotransport response of the YIG/Pt heterostructure for several temperatures, respectively.
        As expected from the SMR theory framework, a $\sin(\alpha)\cos(\alpha)$ modulation is observed in the transverse resistivity.
        The amplitude of this modulation decreases towards higher temperatures and becomes indistinguishable from the background noise and drift above \SI{540}{\kelvin}.
        For the non-local voltage, a $\sin^2(\alpha)$ modulation is observed, where the amplitude increases with temperature until \SI{420}{\kelvin} and then drops off sharply above \SI{500}{\kelvin}.
        Additionally, for $T > \SI{500}{\kelvin}$ a magnetic field orientation independent offset voltage is present in the non-local voltage.
    }
\end{figure}
\twocolumngrid

To study the evolution of the SMR, the MMR and the offset signal as a function of temperature, we used the fits to \eref{smrt} and \eref{nl} to extract the respective amplitudes.
The results are depicted by the symbols in \fref{f2} for the local magnetoresistance (panel (a)) and the non-local magnetoresistance (panel (b)).
Additionally, high temperature magnetization measurements were carried out (full teal line in panel (a)) and compared to a mean-field model for the spontaneous magnetization (dashed orange line) using the parameters put forward by Anderson in Ref.~\cite{Anderson1964}.
The simulation and the measurement show good agreement, highlighting the bulk-like magnetic properties of the LPE-grown YIG films.

In contrast to other reports in Refs.~\cite{Uchida2015, Wang2015} we find a decrease of the SMR linearly with the measured saturation magnetization in our samples. 
Assuming that the dominant scaling of the SMR at high temperatures is determined by the real part of the spin mixing conductance $G_\mr{r} \propto T_\mr{C} - T$ as proposed by recent theories in Refs.~\cite{Zhang2019, Ohnuma2014} and reported experimentally in Refs.~\cite{Uchida2015, Wang2015}, one would expect that the SMR$(T)$ does \emph{not} scale with $M_\mr{s}(T)$, since $M_\mr{s} \propto (T_\mr{C} - T)^{1/2}$ (for a mean field model).\cite{Uchida2015, Chen2013}
We speculate that our findings differ from those in Refs.~\cite{Uchida2015, Wang2015} since we use rotations of the magnetic field instead of magnetic field sweeps to quantify the SMR amplitude. 
In our experiments, the YIG magnetization thus is fully saturated in good approximation, while magnetic domain formation and hysteresis around zero field do not play a role.
The latter could impact the signal shape and thus hamper the quantitative determination of the full SMR magnitude.
An alternative explanation for the observed scaling might be that the SMR in our heterostructures is dominated by the imaginary part of the spin mixing interface conductance $G_\mr{i}$, which would yield a $(T_\mr{C} - T)^{1/2}$ scaling according to theory.\cite{Zhang2019}
However, since typically $G_\mr{i} \ll G_\mr{r}$, this explanation seems unlikely.\cite{Althammer2013, Chen2013, Nakayama2013} 

In addition to the scaling with temperature just discussed, a sharp drop of the SMR is observed already at around \SI{500}{\kelvin}, well below the Curie temperature $T_\mr{C} = \SI{559}{\kelvin}$ of YIG.\cite{Anderson1964}
Please note, that we carried out two experimental runs, where for the first, the maximum temperature was \SI{500}{\kelvin} (gray squares in \fref{f2}), while for the second the maximum temperature was \SI{600}{\kelvin} (red circles) to evaluate possible damage to the sample.
In each run, we started with a magnetic field rotation at \SI{300}{\kelvin} and then ramped up the temperature in steps of \SI{10}{\kelvin}(\SI{15}{\kelvin}) for the first (second) run.
In between the two runs, the temperature was lowered to \SI{300}{\kelvin}.
For the second run, a clear decrease of the SMR is observed already at \SI{300}{\kelvin}.
We speculate that this is related to Fe or Pt interdiffusion at the interface, permanently lowering the interface transparency and thus the SMR magnitude.
Similar annealing effects were reported in Ref.~\cite{Avci2017} in heterostructures of \ch{Tm3Fe5O12} and Pt.
Consequently, the drop and disappearance of the SMR well below $T_\mr{C}$ might be related to a ``dirty'' (intermixed) interfacial layer with a lowered Curie temperature and/or modified spin transparency, since the magnetic properties of the interface critically affect the SMR.\cite{Zhang2019, Velez2019} 
Such an interface layer with modified magnetic properties could also be at the origin of the scaling with temperature discussed above.
Note, that the sample remained at high temperatures for extended periods of time, since the stabilization of the temperature and the angle resolved magnetoresistance measurement require roughly \SI{1.5}{\hour} for each temperature.
Notably, after the run to \SI{600}{\kelvin} (and consequently around $\gtrsim\SI{48}{\hour}$ at elevated temperatures), the sample did not show any SMR whatsoever even at room temperature.

\begin{figure}[h!]
    \includegraphics{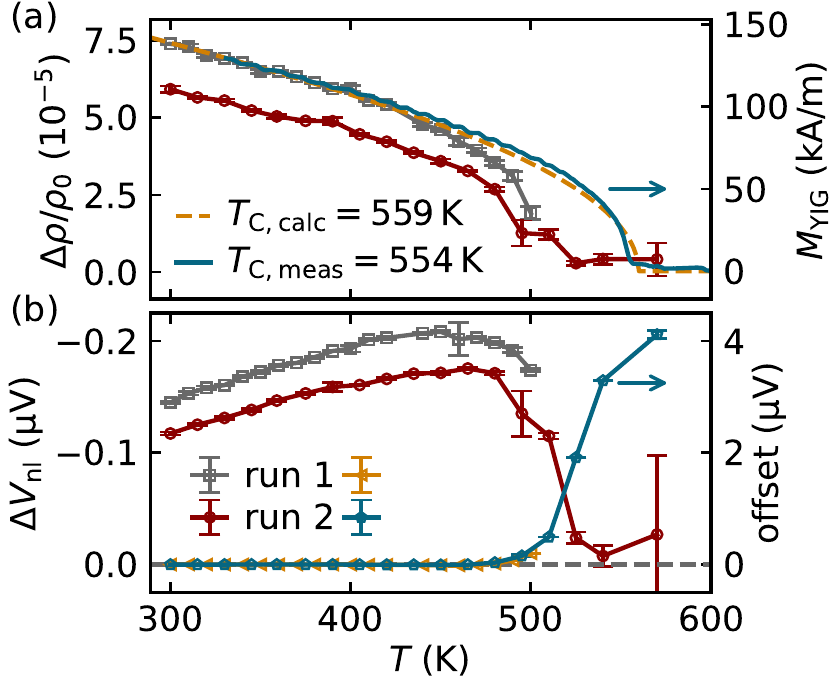}
    \caption{\label{fig:f2}
	(a) SMR amplitude as a function of temperature extracted from $\sin(\alpha)\cos(\alpha)$ fits to the field-orientation dependent measurements (\cf \fref{f1}). 
        The magnetization as a function of temperature of a different piece from the same YIG wafer is shown as solid teal line and the magnetization calculated using a mean-field model is shown as dashed yellow line.
        Until \SI{450}{\kelvin}, the magnetization and the SMR amplitude nicely overlap. However, above \SI{500}{\kelvin} the latter drops off sharply and vanishes well below $T_\mr{C}$ of YIG.
	(b) The amplitude of the non-local voltage modulation $\Delta V_\mr{nl}$ obtained by $\sin^2(\alpha)$ fits to the field-orientation dependent measurements increases from \SI{300}{\kelvin} to \SI{450}{\kelvin} and then drops off sharply.
        As for the SMR, the signal vanishes already well below $T_\mr{C}$.
        In addition, a field-independent offset voltage emerges and drastically increases above $T \approx \SI{490}{\kelvin}$.
        The data shown with gray squares (yellow triangles) were taken during the first measurement run from \SI{300}{\kelvin} to \SI{500}{\kelvin}, the data shown by red circles (teal pentagons) during the second run from \SI{300}{\kelvin} to \SI{600}{\kelvin}.
    } 
\end{figure}

As already seen in \fref{f1}(d), the MMR shows a significantly different trend as a function of temperature (\cf \fref{f2}(b)), increasing towards high tempratures and then dropping off at \SI{500}{\kelvin} similar to the SMR.
The increase with temperature is expected for the MMR, where the magnon generation (and thus $\Delta V_\mr{nl}$) is expected to scale like $T^{3/2}$ at room temperature.\cite{Zhang2012a, Goennenwein2015}
We again associate the sharp drop observed around \SI{500}{\kelvin} to interdiffusion and dirty interface formation, reducing the orbital overlap, and thus the magnon generation efficiency, between the spin Hall active Pt layer and the YIG layer.
Interestingly, both, the SMR and the MMR are decreased by \SI{24}{\percent} from the first to the second run, so that the interface transparency is not necessarily the sole explanation for the amplitude drop.
In particular, the spin Hall angle or the spin diffusion length of the Pt layer might also be altered by Fe interdiffusion, giving rise to the lower SMR and MMR signal amplitudes.

Furthermore, the observed field-orientation independent positive voltage offset in the non-local signal becomes finite above $T\approx\SI{470}{\kelvin}$, increasing drastically for higher temperatures (\cf \fref{f2}(b)).
We attribute this to the onset of electrical conductivity in the YIG layer (or the dirty interface layer), in good agreement with reports using high current densities to locally increase the YIG temperature via ohmic heating.\cite{Thiery2018, Thiery2018a}
More details on the temperature dependence and a model for the mechanism responsible for the non-local offset will be presented below (\cf \fref{f4}).

We now first elucidate the changes that the YIG/Pt heterostructure undergoes at high temperature. 
To that end, the resistivity $\rho_0$ of the Pt layer is shown as a function of $T$ in \fref{f3}(a). 
Until $T\approx\SI{450}{\kelvin}$, $\rho_0$ increases linearly (see gray line) with temperature, while the increase is less pronounced for high temperatures.
For the second run (red symbols), $\rho_0$ is reduced compared to the first run already at \SI{300}{\kelvin}.
We interpret this as evidence for a finite conductivity of the parasitic interface layer (\egg an \ch{FePt} layer).
Consequently, the assumption that either the Pt layer itself or the interface changes, seems well supported by our data.
Please note, that this layer cannot extend over a large area of the heterostructure, since the non-local offset voltage vanishes at \SI{300}{\kelvin} even after annealing.
Consequently, no conductive channel is present at room temperature, so that the changes of the interface must be localized to within at most \SI{500}{\nano\meter} (\ie half the distance between the two Pt electrodes).

\begin{figure}[h!]
    \includegraphics{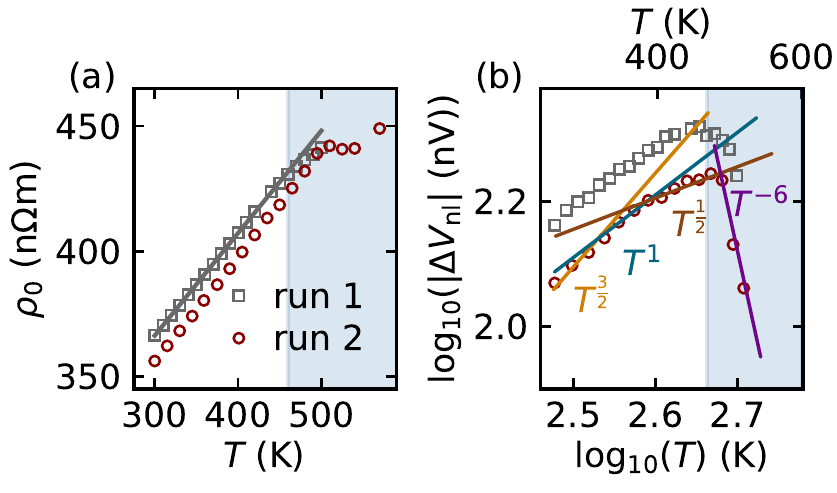}
    \caption{\label{fig:f3}
	(a) The resistivity of the Pt layer increases linearly as a function of temperature until approximately \SI{470}{\kelvin}.
    Above \SI{470}{\kelvin} (blue shaded region), the resistivity increases in a non-linear fashion and remains lower even when the sample is cooled down to \SI{300}{\kelvin} and when the measurement is repeated (red symbols).
	(b) The MMR amplitude $\Delta V_\mr{nl}$ roughly scales as $T^1$ up to $T\approx\SI{470}{\kelvin}$, and then decreases sharply.
    } 
\end{figure}

\Fref{f3}(b) shows the evolution of the MMR signal amplitude $\Delta V_\mr{nl}$ as a function of $T$ in a double logarithmic plot.
Good agreement with a power law $T^\alpha$ behavior reported in similar heterostructures below room temperature is found with $\alpha \approx 1$ for $\SI{300}{\kelvin} < T < \SI{470}{\kelvin}$. 
Zooming in, our data would also be compatible with $\alpha = 3/2$ at \SI{300}{\kelvin},\cite{Goennenwein2015} decreasing gradually with temperature to $\alpha \sim 1/2$ for $T > \SI{400}{\kelvin}$.
We associate this change with the increase of the thermal magnon population, leading to a reduction of the magnon diffusion length towards higher temperatures.
%\begin{equation}
    %\Delta V_\mr{nl} \propto G_\mr{em} G_\mr{me} \exp(-\frac{d_\mr{nl}}{\lambda}
%\end{equation}
Microscopically, magnon-magnon and magnon-phonon scattering becomes more likely, if more magnons are already present in the YIG layer due to the thermal activation.\cite{Cornelissen2016, Bezuglyj2019}
A second explanation for the change of the magnon diffusion length might be rooted in the onset of electrical conductivity, giving rise to additional (electronic) spin losses.
However, the exact separation of the temperature dependence of the magnon generation and diffusion is beyond the scope of this work as sample aging limits the ability to compare different devices on one chip.

\begin{figure}[h!]
    \includegraphics{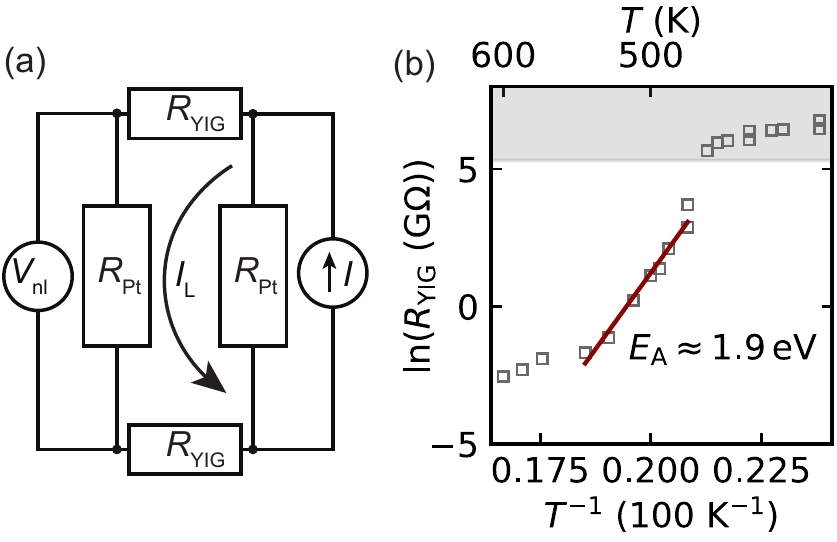}
    \caption{\label{fig:f4}
    \textbf{a} A simplified equivalent circuit of the non-local device allows to evaluate the impact of a finite electrical conductivity of the YIG layer in between the two Pt electrodes.
    If the conductivity of the YIG is non-zero, a small leakage current $I_\mr{L}$ will give rise to an additional magnetic field orientation independent non-local voltage $V_\mr{nl}$.
    \textbf{b} Using this model, the magnetic field-independent offset voltage can be used to determine the resistance $R_\mr{YIG}$ of the YIG film.
    The activation energy of the electrical transport is estimated to $E_\mr{A} \approx \SI{1.9}{\electronvolt}$.
    } 
\end{figure}

Finally, we now address the field-orientation independent offset signal in more detail. 
To that end, we show an equivalent electronic circuit of the device in \fref{f4}(a), where the YIG channel is represented by two resistors with $R_\mr{YIG}$ on the end of the Pt wires ($R_\mr{Pt}$).
For $R_\mr{YIG} = \infty$, the two Pt wires are electrically independent, and a ``clean'' MMR signal can be recorded in the non-local experiment.
If $R_\mr{YIG}$ becomes finite, however, a significant leakage current $I_\mr{L}$ will flow through the YIG channel from one Pt strip to the other, giving rise to an increase of $V_\mr{nl}$ (for equal polarities of $I$ and $V_\mr{nl}$).
Importantly, this leakage current will be independent of magnetic field orientation (assuming that the magnetoresistance of the spurious $R_\mr{YIG}$ is small), and only result in a constant offset voltage.\cite{Thiery2018a}
In turn, this offset voltage can be used to determine $R_\mr{YIG}$ using
\begin{equation}
    R_\mr{YIG} = \frac{I R_\mr{Pt}}{2 V_\mr{nl}} - R_\mr{Pt} \approx \frac{I R_\mr{Pt}}{2 V_\mr{nl}}.\label{eq:ryig}
\end{equation}
Using \eref{ryig} and the field independent offset voltage from both measurement runs (\cf \fref{f2}(b)) we thus estimate $R_\mr{YIG}$ as shown in \fref{f4}(b).
To allow for a better evaluation of the intrinsic scaling behavior we use an Arrhenius representation.
Neglecting the points at very high temperatures (\ie $T > \SI{550}{\kelvin}$), we use $V_\mr{offset} \propto \exp(E_\mr{A}/k_\mr{B} T)$ to estimate the activation energy $E_\mr{A} = E_\mr{g}/2 \sim \SI{1.9}{\electronvolt}$. 
Thus we find the same order of magnitude of the band gap as reported in Ref.~\cite{Thiery2018a} for \SI{18}{\nano\meter} thick LPE grown YIG films $E_\mr{g} \sim \SI{2}{\electronvolt}$ where a Van-der-Pauw measurement is used.
Please note, that the high temperature region is neglected as the resistivity of the Pt layer significantly changes above these temperatures (\cf \fref{f3}(a)), such that it is likely, that the YIG conductivity becomes sizable compared to the Pt layer.
Consequently, the equivalent schematic and \eref{ryig} is not expected to produce reasonable results in this limit (\egg a large part of the current might flow directly from one bond wire to another bond wire instead of through the Pt wire).
Nevertheless, as good agreement is found with the expected band gap, this method is very useful to verify the electronic properties of the magnetic insulator used for non-local transport experiments.

\section{Summary}

In summary, we presented a systematic study of the evolution of the SMR and the MMR in YIG/Pt hetero\-structures for temperatures up to the Curie temperature of YIG.
In particular, we found the SMR to be directly proportional to the saturation magnetization of the YIG layer in our samples, in contrast to recent experimental and theoretical work.
For the MMR we observed a power law dependence on temperature as expected theoretically and motivated by the magnon generation efficiency increasing towards higher temperatures:
However, the power law exponent decreases for higher temperatures, which we attribute to a reduced magnon diffusion length for higher temperatures.
Since more more magnons are thermally excited at elevated temperatures, the reduction of the magnon diffusion length can be understood in terms of scattering of the non-equilibrium, electrically generated magnons from the thermal magnon population, or in terms of electronic spin losses in the YIG layer.
Beyond \SI{470}{\kelvin}, well below the Curie temperature of YIG, both the SMR and MMR decrease drastically and vanish to within our experimental resolution at $T \gtrsim \SI{520}{\kelvin}$.
We speculate that an interfacial layer forms between the YIG and the Pt, permanently reducing the exchange coupling across the interface and possibly altering the SMR scaling.
This sample aging effect makes a detailed investigation of SMR and MMR around and above the Curie temperature $T_\mr{C} \approx \SI{560}{\kelvin}$ of YIG impossible.
Finally, we show that the electronic properties of the YIG film or the interface can be estimated from the non-local offset voltage arising due to leakage currents through the YIG layer in between the two Pt electrodes.
Using this simple model, we find good agreement with the activation energy of the electronic transport reported in literature using a conventional four point measurements.

\acknowledgments

We thank T. Sch{\"o}nherr, M. Lammel and S. Piontek for technical support and A. Erbe and K. Nielsch for access to their research facilities. We acknowledge financial support by the Deutsche Forschungsgemeinschaft via SFB 1143 (project no.\ C08) and through the Würzburg-Dresden Cluster of Excellence on Complexity and Topology in Quantum Matter - ct.qmat (EXC 2147, project-id 39085490).

%\section*{Data Availability}
%
%The data that support the findings of this study are available from the corresponding author upon reasonable request.
%
\bibliography{190618_bibliography.bib}
\end{document}